\documentclass[9pt,sigconf,letterpaper,nonacm]{acmart}

\usepackage{booktabs}
\usepackage{graphicx}
\usepackage[nolist,nohyperlinks]{acronym}
\usepackage{xcolor}
\usepackage{xspace}
\usepackage{siunitx}

\usepackage[capitalise,noabbrev]{cleveref}

\newif\ifdraft
\drafttrue

\ifdraft
  \newcommand{\note}[2]{\textcolor{blue}{\textsf{[\textbf{#1:} #2]}}}
  \newcommand{\missing}[1]{\textcolor{red}{\textsf{[\textbf{PENDING:} #1]}}}
\else
  \newcommand{\note}[2]{}
  \newcommand{\missing}[1]{\PackageError{paper}{Unfilled \string\missing:
    #1}{Fill in the real measured value before building the final PDF.}}
\fi

\newcommand{\eg}{e.g.,\xspace}

\newcommand{\etal}{et al.\xspace}
\newcommand{\rcbg}{Reactive \ac{CBG}\xspace}
\newcommand{\cbg}{\ac{CBG}\xspace}
\newcommand{\vfour}{IPv4\xspace}
\newcommand{\vsix}{IPv6\xspace}

\makeatletter
\AtBeginDocument{\fancyhead[LE,RO]{\@headfootfont\acmConference@shortname,
  \acmConference@date, \acmConference@venue}}
\makeatother

\acmConference[NewGeo '26]{New Directions in IP Geolocation Workshop}{October 12,
  2026}{Karlsruhe, Germany}

\begin{document}

\begin{acronym}
  \acro{AS}{Autonomous System}
  \acrodefplural{AS}[ASes]{Autonomous Systems}
  \acro{ASN}{\ac{AS} Number}
  \acro{AUP}{Acceptable Use Policy}
  \acro{BGP}{Border Gateway Protocol}
  \acro{CBG}{Constraint-Based Geolocation}
  \acro{CCPA}{California Consumer Privacy Act}
  \acro{CDN}{Content Distribution Network}
  \acro{CPE}{Customer Premises Equipment}
  \acro{CSIRT}{Computer Security Incident Response Team}
  \acro{DAD}{Duplicate Address Detection}
  \acro{EUI}{Extended Unique Identifier}
  \acro{GDPR}{General Data Protection Regulation}
  \acro{IoT}{Internet of Things}
  \acro{ISP}{Internet Service Provider}
  \acro{IID}{Interface Identifier}
  \acro{LAN}{Local Area Network}
  \acro{NIC}{Network Interface Card}
  \acro{NTP}{Network Time Protocol}
  \acro{MAC}{Media Access Control}
  \acro{OS}{Operating System}
  \acro{OUI}{Organizationally Unique Identifier}
  \acro{RTT}{Round-Trip Time}
  \acro{SOHO}{Small Office-Home Office}
  \acro{U/L}{Universal/Local}
  \acro{SLAAC}{Stateless Address Autoconfiguration}
  \acro{VPS}{Virtual Private Server}
\end{acronym}

\title[Reactive Constraint-Based Geolocation of Internet Hosts]{Reactive Constraint-Based Geolocation \\ of Internet Hosts}

\author{Spencer Ye}
\email{sye15@jh.edu}
\affiliation{%
  \institution{Johns Hopkins University}
  \city{Baltimore}
  \state{MD}
  \country{USA}
}

\author{Chase Kanipe}
\email{ckanipe1@jhu.edu}
\affiliation{%
  \institution{Johns Hopkins University}
  \city{Baltimore}
  \state{MD}
  \country{USA}
}

\author{Peter Ryan}
\email{pr2427@nyu.edu}
\affiliation{%
  \institution{New York University}
  \city{New York}
  \state{NY}
  \country{USA}
}

\author{Erik Rye}
\email{rye@jhu.edu}
\affiliation{%
  \institution{Johns Hopkins University}
  \city{Baltimore}
  \state{MD}
  \country{USA}
}

\begin{abstract}
Active IP geolocation techniques rely on the responsiveness of Internet hosts,
    while passive techniques depend on data sources that are unevenly adopted and prone
    to staleness and error. In this work, we invert the active IP geolocation
    problem by listening at a geographically distributed set of vantage points
    for unsolicited Internet scans. By \emph{reactively} completing connections
    with these scanners, we obtain round-trip time (RTT) measurements from hosts
    that may otherwise be unresponsive and use those measurements for geolocation.

Over the course of 20 days in August 2026, we recorded \num{46.8} million scans
from \num{289746} distinct scanners, with a reactive RTT for each scan.
We show that reactive RTTs align with those obtained by sending
active probes, with more than three-quarters of \num{990204} measurement
pairs differing by at most \num{10}\,ms. Over half of the scanners that yielded
a 
    reactive RTT did not answer our active probes,
    highlighting the value of our approach. We demonstrate the feasibility
    of multilateration using purely reactive RTTs for \num{56112}
    distinct IP addresses.
Finally, using speed-of-light constraints, we refute commercial
geolocation claims for \num{6852} addresses.
\end{abstract}

\maketitle

\section{Introduction}
\label{sec:intro}

Traditional IP geolocation methods are either active or passive. \emph{Active}
techniques, \eg \cbg~\cite{gueye2004constraint}, send packets toward the
target to induce a response. \emph{Passive} methods estimate a target's location without direct
interaction, \eg from geofeeds or databases.

Active IP geolocation relies on the responsiveness of routers and
hosts to obtain \ac{RTT} measurements, but filtering policies may prevent active
probes from reaching their targets, prevent a response from being generated, or
prevent that response from returning to a vantage point~\cite{bano2018liveness,heidemann2008census}. Passive techniques do not require the target to be
responsive, but frequently require the cooperation of a service provider.
Geofeeds, a popular passive mechanism, are unevenly adopted and may report
stale or incorrect locations.  

In this work, we introduce a third category, \emph{reactive} IP geolocation,
in which the target initiates the interaction. Our technique,
called \rcbg, inverts the well-known \cbg approach. 
Reactive CBG uses unsolicited inbound connections from Internet-wide
scanners~\cite{durumeric2013zmap,durumeric2014scanning,richter2019scanning},
whose location shapes what they observe~\cite{wan2020origin}, rather than
initiating scans on its own.
When these scanners respond to our SYN-ACK with an ACK or RST, we obtain an
\ac{RTT} measurement.

We leverage these unsolicited scans for two geolocation purposes. First, because
refuting a claimed location requires only a single vantage point, we demonstrate
that the \acp{RTT} we receive can be used to falsify IP geolocations of two
widely used commercial geolocation databases. Second, when we obtain RTT measurements
to an Internet-wide scanner from three or more of our vantage points
(whose locations are known), we can multilaterate its position. 

In sum, this work includes the following contributions:

\begin{enumerate}
    \item Twenty-day deployment of a \rcbg platform using \num{42}
        cloud-hosted vantage points in
        \num{29} countries on six continents (\S\ref{sec:results}).
    \item Validation of the consistency between reactive and active RTT
        measurements (\S\ref{sec:validation-results}).
    \item Falsification of the geolocations of \num{6852} addresses in two
        commercial geolocation databases (\S\ref{sec:falsification-results}).
    \item Demonstration of the efficacy and limitations of \rcbg-based localization
        (\S\ref{sec:multilateration-results}).
    \item Public release of our \rcbg system~\cite{rcbg-code}.
\end{enumerate}

\section{Background and Related Work}
\label{sec:background}

Over the last 25 years, researchers have developed various techniques for IP geolocation.
We describe active geolocation, which probes the target;
passive geolocation, which does not interact with the target; and reactive network
telescopes, which respond to unsolicited traffic.

\subsection{Active Geolocation}

In 2001, Padmanabhan and Subramanian published an early empirical study on IP
geolocation. One of the techniques used, GeoPing, established the basis for
active geolocation: ping a target from many geographically distributed landmarks
and infer position from the resulting delay
vectors~\cite{padmanabhan2001investigation}.

\cbg was introduced by Gueye~\etal in 2004. \cbg
converts delays into distance upper bounds and intersects the resulting disks
into a feasible region~\cite{gueye2004constraint}. Katz-Bassett~\etal later
factored router topology into \cbg~\cite{katzbassett2006topology}.

Wong~\etal generalized the constraint framework to include both positive and
negative constraints, respectively representing where the target must and cannot
lie~\cite{wong2007octant}. Youn~\etal reframed geolocation statistically,
replacing hard distance bounds with maximum-likelihood estimates over the
joint distribution of delay and distance~\cite{youn2009statistical}. Wang~\etal and
Rye~\etal developed IP geolocation techniques that incorporate additional,
external ``hints'' about an IP's
location~\cite{wang2011streetlevel,rye2023ipvseeyou}; Darwich~\etal undertook a
replication study of Wang's work~\cite{darwich2023replication}.

\subsection{Passive Geolocation}

There has been significant recent interest in using provider-reported
geolocation information (called ``geofeeds''), since providers ostensibly have
ground truth information about where their prefixes are being
used~\cite{bush2024rfc}. However, uneven adoption, stale or incorrect
geolocation information, and malicious behavior can confound their
utility~\cite{livadariu2024geofeeds,fainchtein2024you}. Commercial geolocation
databases such as MaxMind~\cite{maxmind} and IPinfo~\cite{ipinfo} are now a
widely used source of IP geolocation data, and their (in)accuracy has been the
subject of significant prior
work~\cite{poese2011unreliable,gouel2021stability,gharaibeh2017routergeo}.

\subsection{Reactive Telescopes}

Prior work has used techniques similar to \rcbg, in which a vantage point
responds to traffic the target sends first.
Hiesgen~\etal~\cite{hiesgen2022spoki} operated a reactive network telescope that
answered unsolicited TCP SYNs in real time to elicit the second phase of
stateless scans, characterizing scanning activity invisible to passive
telescopes. Responsive vantage points of this kind date back to
2004~\cite{yegneswaran2004sinks,provos2004virtual,bailey2005internet,pang2004background}.

\section{Methodology}
\label{sec:methodology}

\begin{figure}[t]
  \centering
  \includegraphics[width=\columnwidth]{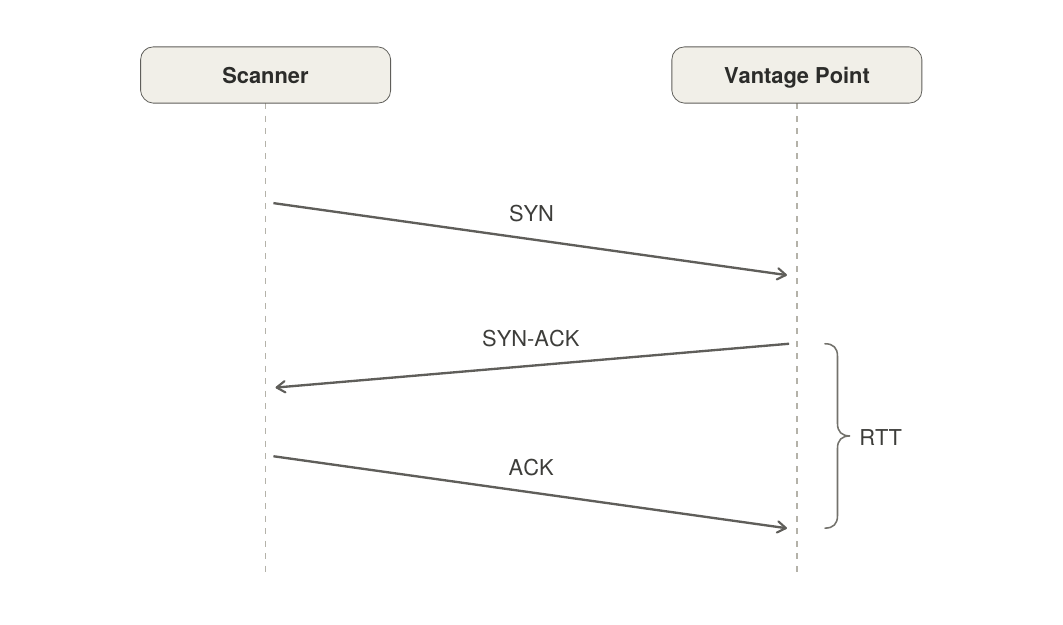}
    \caption{Our system measures the elapsed time between sending a SYN-ACK and
    receiving an ACK or RST from an Internet scanner.}
  \label{fig:tcp-rtt}
\end{figure}

Our system measures the RTTs of unsolicited Internet scans for the purposes of
IP geolocation. To do so, we listen for incoming TCP SYN packets on \num{23} ports,
chosen because they are well-known ports for common services and because they
are frequently subject to Internet-wide port scans and banner grabs
(Table~\ref{tab:ports} in the Appendix lists the ports). 

When an incoming scan occurs, we reply with a standard SYN-ACK packet.
Upon receiving either an ACK or RST in response to our SYN-ACK, we compute the
RTT and store that value for the IP address that scanned our vantage
point~\cite{jiang2002passive}.\footnote{We use the minimum of the RTT computed from \texttt{libpcap}
timestamps and the Linux kernel's \texttt{tcpi\_min\_rtt} value for the
socket.} Figure~\ref{fig:tcp-rtt} shows the elapsed time we compute when the
scanner replies with an ACK; the RST case is equivalent. For each scanner and
vantage point pair, we keep the minimum RTT over every connection during the
capture period and use it for falsification (\S\ref{sec:falsification-methodology})
and multilateration (\S\ref{sec:multilateration}).

\subsection{Reactive RTT Validation Methodology}
\label{sec:reactive-rtt-methodology}

To validate reactive RTTs against active RTTs, each vantage point sends three
active probes to a scanner immediately after recording its reactive RTT.
We send an ICMP (or ICMPv6) Echo Request
and one TCP SYN each to ports 80 and 443. If we receive an ICMP Echo Reply in response
to the Echo Request, or a TCP SYN-ACK or RST in response to the TCP SYNs, we
record the elapsed time as the \emph{active} RTT for that scanner. Not all
scanners respond to our active probes. For example, a firewall may permit this
scanner-initiated exchange while blocking independently initiated probes.
We do not consider ICMP error
messages (\eg port unreachable, administratively prohibited) to be valid
replies. To avoid re-probing the same scanner repeatedly, each vantage point
enforces a 24-hour cooldown before it actively measures an IP address again.

\subsection{Falsification Methodology}
\label{sec:falsification-methodology}

A vantage point that observes a scanner at an RTT of $r$ milliseconds bounds it to
a disk of radius $100r$\,km; data traverses fiber at close to
$\frac{2}{3}c$~\cite{singla2014internet}, so a
millisecond of RTT covers at most \num{100}\,km of separation. We use the full
$\frac{2}{3}c$ bound rather than the tighter $\frac{4}{9}c$ that
\citet{katzbassett2006topology} show holds for all but \num{0.04}\,\% of
measured paths. A claimed location outside of that $100r$\,km disk is
impossible, irrespective of the scanner's true
position~\cite{gharaibeh2017routergeo,weinberg2018proxies}.

We select the vantage point with the smallest RTT for each scanner. We discard
the source if the distance from that vantage point to any other observing vantage
point exceeds the sum of the two RTT-derived distance bounds plus \num{50}\,km.
Otherwise, we test the location claim against the selected vantage point.
A claim is impossible if its distance from that vantage point, less
any accuracy radius the provider publishes, exceeds the bound by a margin of
\num{50}--\num{300}\,km set by the vantage point's positional uncertainty
(Table~\ref{tab:vantage-points} in the Appendix).

\subsection{Multilateration Methodology}
\label{sec:multilateration}

We follow the \ac{CBG}~\cite{gueye2004constraint} methodology. For
each vantage point we fit a \emph{bestline}, the tightest line
$\mathrm{RTT} = a + b\,d$ lying below that vantage point's observed pairs of
distance $d$ and RTT, with intercept $a$ and slope $b$; it converts a measured
RTT $r$ into a distance ceiling $d_{\max} = (r - a)/b$, or the speed-of-light
ceiling $100r$\,km where $r$ falls below the intercept. Each ceiling defines a
disk, and the scanner must lie in their intersection.

We fit bestlines to the RTT mesh among our own vantage points, not to external
landmarks. For each ordered vantage-point pair, we use the minimum
TCP handshake RTT from repeated connections to port 8443.
We intersect geodesic disks by grid search, coarse then fine at a resolution
proportional to the tightest disk, seeding the fine pass from the
least-violating point when the coarse pass misses. We report the region's
area and centroid; sources with no common intersection are left unresolved
rather than relaxing any bound.

\section{Results}
\label{sec:results}

We deployed our \rcbg system to \num{42} vantage points in \num{29} countries
for \num{20} days from August 10--29, 2026.
In this section, we describe the scan corpus, compare reactive and active RTTs,
and present our falsification and multilateration results.

\subsection{Corpus}

We received \num{46.8} million scans. Most (\num{92}\,\%) arrived over \vfour;
because our servers are part of the NTP Pool~\cite{ntppool}, their \vsix
addresses are public and drew some \vsix
scanning~\cite{glowing,richter2022illuminating}.

We received a high volume of scanning, particularly on ports 445 and 5900, from a
small number of scanners; in total we observed \num{289746} unique source IP
addresses (Table~\ref{tab:corpus}). The ratio of \vfour to
\vsix source IPs is about the same as the ratio of scans, with \num{90}\,\% of source
IPs being \vfour addresses.

\begin{table}[t]
  \centering
  \caption{Scans and scanners observed during the experiment.}
  \label{tab:corpus}
  \footnotesize
  \setlength{\tabcolsep}{4pt}
  \begin{tabular}{lrrrr}
    \toprule
    & \vfour & \vsix & Total & \% \\
    \midrule
    Scans seen                   & \num{43174815} & \num{3615196} & \num{46790011} & --- \\
    \addlinespace[2pt]
    Unique addresses               & \num{261875} & \num{27871} & \num{289746} & 100.0 \\
    \addlinespace[2pt]
    \quad no active RTT            & \num{145701} & \num{8017}  & \num{153718} &  53.1 \\
    \quad seen by 1 vantage point  & \num{173830} & \num{16312} & \num{190142} &  65.6 \\
    \quad seen by $\geq$\,3        & \num{54928}  & \num{8002}  & \num{62930}  &  21.7 \\
    \quad tested ($\leq$\,\num{100}\,ms) & \num{93796} & \num{19859} & \num{113655} & 39.2 \\
    \quad falsified                & \num{6503}   & \num{349}   & \num{6852}   &   2.4 \\
    \bottomrule
  \end{tabular}
\end{table}

\begin{figure}[t]
  \centering
  \includegraphics[width=\columnwidth]{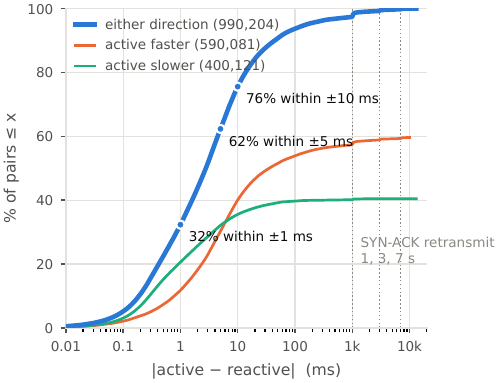}
    \caption{CDF of the absolute difference between the reactive and active
    RTTs measured to the same scanner, on a logarithmic axis, for all pairs and
    split by which RTT was smaller. Dotted lines mark the SYN-ACK
    retransmission timers at \num{1}, \num{3}, and
    \num{7}\,s~\cite{paxson2011rfc}.}
  \label{fig:rtt-signed}
\end{figure}

\begin{table}[t]
  \centering
  \caption{Source IP addresses by the organization type of their Autonomous System owners.}
  \label{tab:as}
  \footnotesize
  \setlength{\tabcolsep}{4pt}
  \begin{tabular}{lrrrr}
    \toprule
    Category & Count & \% \\
    \midrule
    Internet Service Provider      & \num{207363} & 71.56 \\
    Hosting                        & \num{77543} & 26.76 \\
    Business      		   & \num{3649} & 1.25 \\
    Education			   & \num{572} & 0.19 \\
    Unknown			   & \num{340} & 0.11 \\
    Government                     & \num{279} & 0.09 \\
    \addlinespace[2pt]
    \bottomrule
  \end{tabular}
\end{table}

Table~\ref{tab:as} provides a high-level categorization of the organizations
listed as registered owners of the Autonomous Systems contained within the
corpus, binned per IPinfo's IP to Company Database~\cite{ipinfo}. The vast
majority of source IP addresses ($>$98\,\%) were associated with organizations
classified as ``Internet Service
Provider'' or ``Hosting'', with cloud service providers accounting for four of
the six organizations with the most unique source IP addresses.

\subsection{Reactive RTT Validation Results}
\label{sec:validation-results}

To validate that reactive RTTs closely approximate active RTTs, we sent three
types of active probes to every scanner, subject to the cooldown of
\S\ref{sec:reactive-rtt-methodology}. When we obtained multiple active RTTs from
a target, we selected the minimum.

Figure~\ref{fig:rtt-signed} displays the difference between the reactive and
active RTTs over \num{990204} measurement pairs. The majority (\num{75.5}\,\%) of
RTT differences are \num{10}\,ms or less, with a median absolute difference of
\num{2.8}\,ms. In relative terms, \num{78}\,\% of RTT pairs differ by at most
\num{10}\,\% of the reactive RTT. The active RTT was faster in \num{590081} pairs
(\num{59.6}\,\%) and slower in \num{400121}  pairs (\num{40.4}\,\%); two pairs
had equal RTTs. 

When reactive RTTs exceed their active analogs, they provide a looser bound for
geolocation rather than producing incorrect falsifications or location estimates.
However, the reactive RTT yields the tighter distance bound in
\num{40.4}\,\% of pairs; since any measured RTT upper-bounds distance, these
tighter bounds remain valid.   Reactive RTTs also provide value when they are
the \emph{only} RTT available, as for \num{153718} of \num{289746}  scanning IP
addresses (\num{53.1}\,\%; Table~\ref{tab:corpus}). Reactive RTTs therefore provide a defensible basis for CBG.

\begin{figure}[t]
  \centering
  \includegraphics[width=\columnwidth]{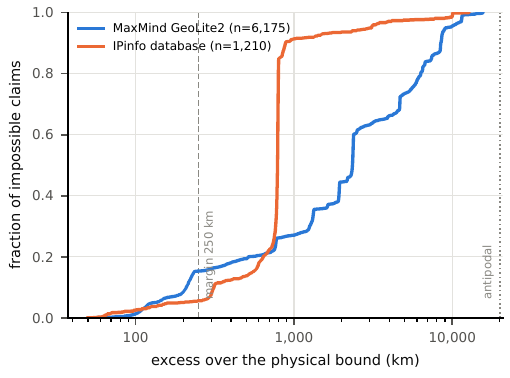}
  \caption{CDF of excess distance beyond our reactive RTT bounds for refuted
    MaxMind and IPinfo location claims. We tested \num{113655} source addresses
    with a reactive RTT of at most \num{100}\,ms.}
  \label{fig:falsification}
\end{figure}

\subsection{Falsification Results}
\label{sec:falsification-results}

Next, we compared IP geolocation data from two popular commercial vendors
(IPinfo~\cite{ipinfo} and MaxMind~\cite{maxmind}\footnote{We used the GeoLite2
City database and researcher access to IPinfo.}) with our
computed reactive RTTs to determine whether any of the reported IP locations can
be falsified. 

First, we filtered scanning IPs for those that had a reactive RTT of $\leq$
\num{100}\,ms. We applied this filter because a \num{100}\,ms RTT creates a disk
with a radius of \num{10000}\,km, covering about half of the Earth's surface. This
resulted in a set of \num{113655} IPs that we then geolocated using IPinfo and
MaxMind. Then, we computed how many of these IP addresses were placed farther from
a vantage point than the maximum distance our reactive RTT allows, with an
additional margin of \num{50}--\num{300}\,km for the vantage point location.
We also subtracted the accuracy radius MaxMind publishes with each entry.

We found \num{6852} (\num{6}\,\%) of these listed at impossible locations:
\num{6175} by MaxMind and \num{1210} by IPinfo, with \num{533} impossible in
both. Of those \num{6852}, \num{2837} (\num{41}\,\%) never answered our active
probes, so two of every five refutations \rcbg produces were unavailable to
our active measurements.  Raising the margin to a uniform \num{250}\,km retains
\num{5227} of MaxMind's refutations and \num{1142} of IPinfo's (\num{85} and
\num{94}\,\%), so the result does not hinge on the per-vantage-point margins.

For each target IP that MaxMind or IPinfo places at an impossible geolocation,
we computed the excess distance beyond what physical bounds permit.
Figure~\ref{fig:falsification} is a CDF of that excess. IPinfo's curve is
dominated by three /24 networks roughly \num{800}\,km beyond the distance the
speed of light permits given our reactive RTTs.

\subsection{Multilateration Results}
\label{sec:multilateration-results}

Only \num{62930} (\num{21.7}\,\%) of scanners connected to three or more
vantage points, the minimum for multilateration.

\begin{figure}[t]
  \centering
  \includegraphics[width=\columnwidth]{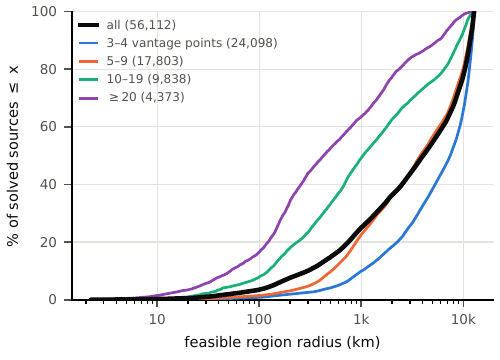}
  \caption{Equivalent radius (that of a disk with the region's area) of the
    feasible region for the \num{56112} of \num{62930} scanners seen by three
    or more vantage points whose constraints had a common intersection. The
    region tightens as more vantage points observe a scanner.}
  \label{fig:locate-cdf}
\end{figure}

Figure~\ref{fig:locate-cdf} shows the size of the feasible region for
\num{56112} IPs whose RTT-derived constraints had a common intersection.
The remaining \num{6818} (\num{11}\,\%) had no common intersection, and their
share rises with the number of constraints, from \num{7}\,\% at three or four
vantage points to \num{36}\,\% at twenty or more. A hard-constraint method fails
whenever one bound is too tight, typically because a scanner's path is faster than any in
our \num{42}-point calibration mesh or the address is anycast; we report them
rather than loosen the bounds, so the medians below cover the resolved subset.

The median feasible region was smaller for scanners observed by more vantage
points. Among the \num{24098} IPs observed by three or four vantage points,
the median equivalent radius was \num{7442}\,km, and only \num{2}\,\% had a
radius of \num{250}\,km or smaller. For the \num{9838} IPs observed by
\num{10}--\num{19} vantage points, the median was \num{1039}\,km, with
\num{2036} (\num{21}\,\%) at or below \num{250}\,km. For the \num{4373} IPs
observed by at least \num{20} vantage points, the median was \num{427}\,km,
with \num{1723} (\num{39}\,\%) at or below \num{250}\,km.

\section{Limitations}
\label{sec:limitations}

Our \rcbg methodology is subject to a number of limitations. First and foremost,
due to the reactive nature of our \cbg implementation, we can only
geolocate hosts that initiate connections to our vantage points, which in
practice limits our targets to Internet-wide scanners. For multilateration, we
require a scanner to initiate a connection to at least three of our vantage
points. CGNAT, last-mile and satellite delays loosen distance bounds, which
affects localization accuracy.
Finally, scanners may scan any TCP port, but we listened on only the \num{23}
ports listed in Table~\ref{tab:ports} in the Appendix.

\section{Conclusion}
\label{sec:conclusion}

In this work, we demonstrated that unsolicited connections from Internet scanners
can be used for geolocation, a technique we call \rcbg. Over
\num{20} days across \num{42} vantage points in \num{29} countries, we recorded \num{46.8} million scans from
\num{289746} distinct sources. From these scans, we obtained reactive RTTs,
which we used to (i) falsify IP geolocations given by two commercial
geolocation services and (ii) coarsely multilaterate scanners that contact at
least three vantage points.
 
Because our reactive measurement system awaits connection attempts from Internet
scanners, it sends no unsolicited traffic, works against targets that do
not answer active probes, and can share infrastructure that provides other
services without interference.
Our results suggest that the primary contribution of \rcbg is
falsification rather than localization, although our localization bounds
tighten as more vantage points observe a scanner. We are open-sourcing our \rcbg
platform~\cite{rcbg-code} so that other researchers can build on our results.

\begin{acks}
We thank the anonymous reviewers and Robert Beverly for their thoughtful feedback.
\end{acks}

\bibliographystyle{ACM-Reference-Format}
\bibliography{conferences,refs}

\clearpage
\appendix

\twocolumn[%
  \begin{center}
    \captionof{table}{The \num{42} vantage points used in this study.
    Margin is the distance by which a
    claim must exceed the speed-of-light bound before we call it
    impossible (\S\ref{sec:falsification-methodology}): the greater of a
    \num{50}\,km floor and that position's own uncertainty, which is
    \num{25}\,km for locations reported in RFC~8805 geofeeds~\cite{kline2020rfc8805}
    or supplied directly by hosting providers,
    \num{100}\,km where we corrected a position by round-trip time, and
    \num{250} to \num{300}\,km for a facility inferred from published
    datacenter or PeeringDB records.}
    \label{tab:vantage-points}
    \footnotesize
    \setlength{\tabcolsep}{3pt}
  \begin{tabular}{llr c llr c llr c llr}
    \toprule
    City & Ctry. & Margin & & City & Ctry. & Margin & & City & Ctry. & Margin & & City & Ctry. & Margin \\
    \midrule
    Dubai & AE & 50 & & Aubervilliers & FR & 50 & & Oslo & NO & 50 & & Dallas & US & 50 \\
    Port Melbourne & AU & 50 & & Worcester & GB & 250 & & Warsaw & PL & 50 & & Elk Grove Village & US & 50 \\
    Sofia & BG & 50 & & Hong Kong & HK & 50 & & Belgrade & RS & 50 & & Honolulu & US & 50 \\
    Osasco & BR & 50 & & Budapest & HU & 50 & & Moscow & RU & 50 & & Kent & US & 50 \\
    Montréal & CA & 300 & & Rosh HaAyin & IL & 50 & & Spånga & SE & 50 & & Lenexa & US & 250 \\
    Santiago & CL & 50 & & Mumbai & IN & 50 & & Singapore & SG & 100 & & Los Angeles & US & 50 \\
    Limassol & CY & 50 & & Noida & IN & 50 & & Singapore & SG & 100 & & Miami & US & 50 \\
    Baden-Baden & DE & 50 & & Tokyo & JP & 50 & & Singapore & SG & 50 & & Piscataway & US & 50 \\
    Berlin & DE & 250 & & Seoul & KR & 50 & & Istanbul & TR & 50 & & Isando & ZA & 50 \\
    Tallinn & EE & 50 & & Almaty & KZ & 50 & & Atlanta & US & 50 & &  &  &  \\
    Logroño & ES & 250 & & Santiago de Querétaro & MX & 50 & & Dallas & US & 50 & &  &  &  \\
    \bottomrule
  \end{tabular}
  \end{center}
  \par\addvspace{0.5\baselineskip}
]
 \section*{Appendix}
\label{app:vantage-points}

\subsection*{Ethical Considerations}
\label{sec:ethics}

Our reactive measurements do not initiate traffic. Instead, we respond only to
unsolicited TCP SYNs we receive from Internet scanners. We reply to these
scans with only a SYN-ACK packet to derive an RTT from a subsequent TCP ACK or
RST. 

When we obtain a reactive RTT for an Internet scanner, we conduct follow-up
measurements to compute an active RTT. These probes are sent only to an address
that has just connected to us and consist of one ICMP Echo Request and one TCP
SYN to each of ports \num{80} and \num{443}. Therefore, for each scan that is
directed at one of our vantage points, we send at most one SYN-ACK and three
probes.

To prevent excessive probing, we enforce a \num{24}-hour cooldown between active
probe rounds to the same address from the same vantage point, along with a
sustained rate limit of two probes per second. Over the capture period we conducted \num{1.46} million
probe rounds, about \num{1700} per vantage point per day, a small fraction of
the traffic the probed hosts had initiated to our vantage points. We received no
abuse reports or complaints during our study. We believe these limits are
sufficient to prevent harm to scanners that contact our vantage points and are
consistent with the principles of the Menlo Report~\cite{menlo}.

\subsection*{Vantage Points and TCP Ports Used}

Table~\ref{tab:vantage-points} lists the cloud-hosted vantage points we used
for our \rcbg measurements. Table~\ref{tab:ports} lists the
\num{23} TCP ports on which we responded to unsolicited incoming scans to
obtain a reactive RTT measurement.

\begin{table}[!h]
  \centering
    \caption{The \num{23} TCP destination ports our vantage points accepted
    unsolicited connections on, with each port's share of the \num{46.8} million
    scans and of the \num{289746} distinct scanners. Scanner percentages total
    more than \num{100}\,\% because a single scanner frequently probed several
    ports.}
  \label{tab:ports}
  \footnotesize
  \setlength{\tabcolsep}{3.5pt}
  \begin{tabular}{rlrrrr}
    \toprule
    & & \multicolumn{2}{c}{Scans} & \multicolumn{2}{c}{Scanners} \\
    \cmidrule(lr){3-4}\cmidrule(lr){5-6}
    Port & Service & count & \% & count & \% \\
    \midrule
    \multicolumn{6}{l}{\emph{Web / proxy}} \\
    443 & HTTPS & \num{2154815} & 4.61 & \num{75819} & 26.2 \\
    80 & HTTP & \num{2069905} & 4.42 & \num{73878} & 25.5 \\
    8443 & HTTPS alt & \num{406968} & 0.87 & \num{16231} & 5.6 \\
    8080 & HTTP alt & \num{340945} & 0.73 & \num{30198} & 10.4 \\
    8000 & HTTP alt & \num{249055} & 0.53 & \num{10431} & 3.6 \\
    8888 & HTTP alt & \num{241308} & 0.52 & \num{11275} & 3.9 \\
    3128 & Squid proxy & \num{78809} & 0.17 & \num{6454} & 2.2 \\
    \midrule
    \multicolumn{6}{l}{\emph{Remote access}} \\
    5900 & VNC & \num{22068644} & 47.17 & \num{8740} & 3.0 \\
    23 & Telnet & \num{2921211} & 6.24 & \num{143549} & 49.5 \\
    3389 & RDP & \num{1415173} & 3.02 & \num{37033} & 12.8 \\
    5555 & ADB / misc & \num{104242} & 0.22 & \num{8384} & 2.9 \\
    2323 & Telnet alt & \num{60107} & 0.13 & \num{6204} & 2.1 \\
    \midrule
    \multicolumn{6}{l}{\emph{File transfer / sharing}} \\
    445 & SMB & \num{11485412} & 24.55 & \num{36821} & 12.7 \\
    21 & FTP & \num{537324} & 1.15 & \num{34632} & 12.0 \\
    \midrule
    \multicolumn{6}{l}{\emph{Databases}} \\
    1433 & MSSQL & \num{633460} & 1.35 & \num{36798} & 12.7 \\
    27017 & MongoDB & \num{545841} & 1.17 & \num{10923} & 3.8 \\
    3306 & MySQL & \num{442231} & 0.95 & \num{34008} & 11.7 \\
    9200 & Elasticsearch & \num{302301} & 0.65 & \num{10437} & 3.6 \\
    5432 & PostgreSQL & \num{291396} & 0.62 & \num{9382} & 3.2 \\
    6379 & Redis & \num{142377} & 0.30 & \num{9830} & 3.4 \\
    \midrule
    \multicolumn{6}{l}{\emph{Infrastructure / IoT}} \\
    7547 & TR-069 & \num{106342} & 0.23 & \num{14613} & 5.0 \\
    2375 & Docker API & \num{100831} & 0.22 & \num{5015} & 1.7 \\
    5060 & SIP & \num{91314} & 0.20 & \num{6583} & 2.3 \\
    \bottomrule
  \end{tabular}
\end{table}

\subsection*{Generative AI}

The authors used Anthropic's Claude Opus 5 for assistance with the development
of the \rcbg RTT collection system and analysis tooling.

\clearpage

\end{document}